\documentclass{article}
\usepackage[utf8]{inputenc}
\usepackage{braket}
\usepackage{amsmath}
\usepackage{graphicx}
\usepackage{enumitem}   
\usepackage{mathrsfs}
\usepackage{dsfont}
\usepackage{mathtools}
\usepackage{amssymb}
\usepackage{xcolor}
\usepackage{physics}
\usepackage{comment}
\usepackage{hyperref}
\usepackage{authblk}
\usepackage{cite}
\usepackage{mdframed}
\usepackage{url}

\DeclareMathAlphabet{\pazocal}{OMS}{zplm}{m}{n}

\newcommand{\zerodel}{.\kern-\nulldelimiterspace}

\renewcommand\bra[1]{{\left<#1\right|}}
\renewcommand\ket[1]{{\left|#1\right>}}

\title{Restoring Locality: \\
The Heisenberg Picture as a Separable Description of Quantum Theory}
\author[1]{Sam Kuypers\thanks{samuelkuypers@gmail.com}}
\affil[1]{\normalsize{Département d'informatique et de recherche opérationnelle,} \linebreak
\normalsize{Université de Montréal, Canada}}

\date{\normalsize{\today} \\ Forthcoming in Alyssa Ney (ed.), \textit{Local Quantum Mechanics: Everett, Many Worlds, and Reality}.
New York: Oxford University Press.}

\begin{document}
\maketitle

\begin{abstract}
Local realism has been the subject of much discussion in modern physics, partly because our deepest theories of physics appear to contradict one another in regard to whether reality is local. According to general relativity, it is, as physical quantities (perceptible or not) in two spacelike separated regions cannot affect one another. Yet, in quantum theory, it has traditionally been thought that local realism cannot hold and that such effects do occur. This apparent discrepancy between the two theories is resolved by Everettian quantum theory, as first proven by Deutsch \& Hayden \cite{deutsch2000information}. In this paper, I will explain how local realism is respected in quantum theory and review the advances in our understanding of locality since Deutsch \& Hayden’s work, including the concept of local branching and the more general analysis by Raymond-Robichaud \cite{raymond2021local}.

\end{abstract}

\section{Introduction} \label{introduction}
Ever since Bell's inequalities were tested experimentally, much of the scientific community has concluded that local realism must be false: it is now widely believed that nature allows one system to affect the state of another regardless of the distance between them, albeit in ways that are necessarily imperceptible. The empirical violations of these inequalities are often thought to be in opposition to a view held by Einstein, that the world must be local. As one of the 2022 Nobel Laureates in Physics, John Francis Clauser, said of his empirical demonstration of the violations of Bell's inequalities \cite{McDonald2022}, 
`I was very sad because I thought Einstein was right but, unfortunately, I did disprove his point of view.' 

It is true that the experiments of Clauser and others \cite{freedman1972experimental} refuted local hidden-variable models, which posit that correlations between spacelike-separated systems arise from their shared dependence on a parameter \(\lambda\).\footnote{As Bell notes \cite{bell1964einstein}, it makes no difference to hidden-variable theories whether \(\lambda\) is a set of parameters, a function, or even a set of functions, nor whether it is discrete or continuous.} However, Einstein’s conception of locality was broader. He required that a local theory satisfy two principles:
\begin{itemize}
    \item \textbf{Separability}: The description of a composite system must decompose into the descriptions of its subsystems, and conversely, the composite system is completely specified by the combined descriptions of its subsystems \cite{einstein1948quanten, howard1985einstein, howard1989holism}.\footnote{Howard \cite{howard1985einstein} convincingly argues that Einstein’s reasoning in Ref.~\cite{einstein1948quanten} amounts to the requirement of separability.} A description of a system that satisfies separability will be called \textit{separable}.
    
    \item \textbf{No action at a distance}: The separable description of a system must be independent of operations performed on another system that is dynamically isolated from the first.
\end{itemize}
Note that the no-action-at-a-distance principle relies fundamentally on separability because it should be the separable description of a system which remains unchanged by operations on a second system that is dynamically isolated from the first. Some non-separable descriptions, like density matrices, can also remain unaffected by such operations -- this is a separate condition known as \textit{no-signalling}. The no-action-at-a-distance requirement was neatly summarised by Einstein in the following quote \cite{einstein1949albert}, 
\begin{quote}
\begin{center}
\textit{`On one supposition we should, in my opinion, absolutely hold fast: the real factual situation of system $S_2$ is independent of what is done with $S_1$, which is spatially separated from the former.'}
\end{center}
\end{quote}

Bell locality and local realism in Einstein's sense are distinct concepts. In particular, a local-realistic theory in Einstein's sense can violate Bell's inequalities, as I will explain in more detail in Sec.~\ref{sec:local-branching}. The fact that nature violates Bell's inequalities does not, therefore, imply that Einsteinian local realism is false. As a glimpse of what is ahead, one prominent example of a local-realistic theory that violates Bell's inequalities is unitary quantum theory, since it satisfies both separability and no action at a distance. This was first proven by Deutsch \& Hayden \cite{deutsch2000information}, who presented their results over a quarter of a century ago. Their approach, in essence, is simply to present unitary quantum theory in the Heisenberg picture; the Deutsch–Hayden construction, along with subsequent developments by myself and others (see, for instance, Refs.~\cite{kuypers2021everettian, bedard2021cost, bedard2021abc, raymond2017equivalence, raymond2021local}), will be the centrepiece of this article, as it resolves the apparent conflict between quantum theory and local realism.

\subsection{The objection of Einstein, Podolsky and Rosen} \label{EPR}
Einstein's conviction about local realism is borne out by general relativity, as operations in a spacetime region \(R_1\) cannot influence the description of another region \(R_2\) that is spacelike-separated from the former. For instance, if a mass is moved in \(R_1\), producing effects like gravitational waves, these do not affect in any way the description of \(R_2\), since no such effect can propagate to the spacelike-separated region \(R_2\) \cite{thorne2000gravitation}. So, when \(R_1\) and  \(R_2\) are spacelike-separated, they cannot affect one another, either perceptibly or otherwise. In fact, a key concern about quantum theory violating local realism is that it could undermine the locality of general relativity: since spacetime in general relativity is shaped by the matter inside it, if matter does not adhere to locality, then \textit{prima facie} neither does spacetime.

The concern that quantum theory is non-local stems from its formalism. In the Schrödinger picture, a pure-state system quantum system at a particular instant is described by a state vector, \(\ket{\Psi}\), which is a ray in the Hilbert space of the system. The dynamical evolution of a system maps its state vector at some initial time to another one in the same Hilbert space that represents the system at a later time. 

There are two dynamical processes in conventional quantum theory. Firstly, there is the evolution of isolated systems. These systems evolve unitarily: if the initial state of the system is \(\ket{\Psi}\), then the state of the system at a later time is \(U\ket{\psi}\), where \(U\) is a unitary operator on the Hilbert space of the system. Unitary evolution is deterministic and reversible.

Conventional quantum theory also requires a second, much more problematic dynamical process called ‘collapse’, which occurs during measurement. Suppose a system is initially in state \(\ket{\Psi}\) in  \(\mathcal{H}_{AB}\), describing a composite system with subsystems \(A\) and \(B\). The Hilbert space of the composite system is thus a tensor product of those of \(A\) and \(B\), i.e. \(\mathcal{H}_{AB} = \mathcal{H}_A \otimes \mathcal{H}_B\). If one measures an observable \(\hat{A}\) acting only on subsystem \(A\), then one selects a measurement basis \(\{\ket{a_1}, \ket{a_2}, \dots, \ket{a_n}\}\), where each \(\ket{a_i}\) is an eigenstate of \(\hat{A}\) with corresponding eigenvalue \(a_i\), and lies in the Hilbert space \(\mathcal{H}_A\). In this case, the measurement induces a collapse of the total state as follows:
\begin{equation} \label{eq:collapse}
\ket{\Psi} \overset{\text{collapse}}{\xlongrightarrow{\hspace{1cm}}} \frac{\ket{a_i}\langle {a_i} | \Psi \rangle }{|\langle a_i | \Psi \rangle|}.
\end{equation}
Here, \(\langle a_i | \Psi \rangle\) denotes a \emph{partial} inner product over \(\mathcal{H}_A\); the result is generally an (unnormalised) state in \(\mathcal{H}_B\). When \(\ket{\Psi}\) belongs to the same Hilbert space as \(\ket{a_i}\), then \(\langle a_i | \Psi \rangle\) is simply a complex number.

The particular eigenstate that the state vector collapses into is ‘chosen’ randomly, such that collapse into an eigenstate \(\ket{a_i}\) occurs with probability \(|\langle a_i | \Psi \rangle|^2\). Thus, collapse is a probabilistic process, meaning that, unlike unitary evolution, it is neither deterministic nor reversible.\footnote{Here, I have described collapse in terms of projective measurements, which is not the most general type of measurement. However, projective measurements together with unitary dynamics are sufficient to implement general measurements, as shown in Chapter~2 of Ref.~\cite{nielsen2010quantum}.}

Collapse affects the entire state vector \(\ket{\Psi}\), including those parts that represent subsystems that are not being measured. This is problematic in cases where the state vector is an entangled state because it implies that a measurement-induced collapse of one system will affect the state of a remote system that it is entangled with. Einstein was particularly concerned about this `spooky action'. One place he expressed this concern is in his renowned paper with Boris Podolsky and Nathan Rosen (commonly abbreviated as EPR) \cite{einstein1935can}; there, they discussed a scenario involving two entangled particles and concluded that quantum theory (in the form it existed then) could not be a complete description of physical reality, although they did consider it `correct', meaning it gave the right predictions. The version of the argument I present here follows Einstein’s exposition in his 1948 article \textit{Quanten-mechanik und wirklichkeit} \cite{einstein1948quanten}. That clarifies the implicit criterion in his original 1935 EPR paper, but the core intention is the same.

The EPR argument can be expressed in a modern, quantum-information-theoretic form as follows. Suppose there are two agents, Alice and Bob, who are each in possession of a qubit, namely \(\mathcal{Q}_A\) and \(\mathcal{Q}_B\), respectively, such that the composite system of their two qubits is in the following entangled state
\begin{equation} \label{eq:EPR-pair}
\ket{\Psi}_{AB}=\frac{1}{\sqrt{2}} \bigl( \ket{1}_A \ket{1}_B + \ket{0}_A \ket{0}_B \bigr).
\end{equation}
Here, the subscripts \(A\) and \(B\) denote whether the vector describes Alice's or Bob's qubit. Suppose also that Alice and Bob are spacelike-separated from one another. If Alice measures her qubit, \(\mathcal{Q}_A\), in the computational basis \(\{\ket{0}_A, \ket{1}_A\}\), the post-measurement state vector of the total system collapses into a product state. In particular, due to Alice's measurement, the composite system collapses either into the product state \(\ket{0}_A\ket{0}_B\) or \(\ket{1}_A\ket{1}_B\).

Could it be that the two qubits have a separable description in terms of the state vector? What EPR argued is that if such a description exists, then this separable description of Bob's qubit depends on Alice's measurement of her qubit. The EPR argument can be summarised as follows:
\begin{enumerate}
    \item Assume that there is a separable description of Alice and Bob's qubits in terms of the state vector and that there is no action at a distance. In particular, the no-action-at-a-distance principle implies that a separable description of \(\mathcal{Q}_B\) should be unaffected by Alice's measurement of \(\mathcal{Q}_A\). 
    \item If Alice measures her qubit in the computational basis, then the outcome \(\ket{0}_A\) implies that Bob’s qubit is in state \(\ket{0}_B\), while the outcome \(\ket{1}_A\) implies that Bob’s qubit is in state \(\ket{1}_B\). 
    \item Depending on the outcome of Alice's measurement, Bob's qubit is either in state \(\ket{0}_B\) or \(\ket{1}_B\). Since these two states are mutually exclusive, this implies action at a distance.
\end{enumerate}
In Appendix \ref{sec:noumenal-EPR}, the EPR paradox is formulated more precisely, using the mathematical results of Sec.~\ref{sec:noumenal}.

It follows from the arguments presented above that if one assumes there exists a separable description of the systems in terms of the state vector, then that separable description of Bob's qubit must be affected by Alice's measurement of her qubit. That is, an operation on \(\mathcal{Q}_A\) affects the separable description of \(\mathcal{Q}_B\), violating the no-action-at-a-distance principle. Hence, either no local description exists or that local description is not provided by the state vector, in which case the state vector would be, in the words of EPR, an `incomplete' description.

In hindsight, the EPR paradox highlights two key concerns. The first is that the state vector does not provide a separable description since the state vector of the composite system cannot be decomposed into parts that recombine to the whole. The second problem is caused by the collapse process, which affects the entire state vector, including that part of it describing systems that are not involved in a measurement, such as \(\mathcal{Q}_B\) above, implying action at a distance.

I shall argue that EPR were right to consider quantum theory incomplete (in the form it existed in their day). Completing the theory begins with discarding the collapse postulate, which contributes to the problem described above. This option was not available to EPR because Everett had not yet published his relative state formalism (see Sec.~\ref{sec:everett}), in which there is fundamentally no collapse -- only unitary evolution. Everett’s unitary quantum theory obviates many of the issues of non-locality by sidestepping the problems associated with that postulate.

Within unitary quantum theory, the objection of EPR \textit{can} be addressed because unitary quantum theory permits a local description. Specifically, the Heisenberg picture of unitary quantum theory is both separable and has no action at a distance, as I shall explain through the rest of this article – this local description is one Einstein, Podolsky, and Rosen might well have found satisfactory.

\section{Everettian quantum theory} \label{sec:everett}
One of the key difficulties in providing a local account of quantum theory is the collapse postulate, because a measurement of one system makes the entire state vector collapse, which generally seems to affect the state of remote systems. Collapse, therefore, poses a problem to local realism; this problem is solved by Everett's relative state construction \cite{everett1973theory}. In Everett's construction, the collapse postulate is not assumed to be true, so systems evolve unitarily, even during measurement. Remarkably, unitary evolution can reproduce all the results that are usually derived from the collapse postulate.

The unitary theory of measurement requires that the measuring instrument (or measurer for short) is a quantum system. Consider again a system with an observable \(\hat{A}\) and a set of orthonormal states \(\{\ket{a_1},\ket{a_2},\dots, \ket{a_n}\}\) that are the eigenstates of \(\hat{A}\). Also consider a second system that is a measurer of \(\hat{A}\), which has a set of orthonormal states \(\{\ket{M_{a_1}},\ket{M_{a_2}}, \dots, \ket{M_{a_n}}\}\). By virtue of being orthonormal, these states of the measurer are distinguishable from one another. A general measurement protocol for measuring the observable \(\hat{A}\) is such that for all \(i \in \{1, ..., n\}\), the measurer and the system evolve as follows:
\begin{equation} \label{eq:measurement}
    \ket{M_0} \ket{a_i} \overset{\text{measurement}}{\xlongrightarrow{\hspace{1cm}}} \ket{M_{a_i}} \ket{a_i},
\end{equation}
where \(\ket{M_0}\) is some receptive state of the measurer, not necessarily distinct from any of the elements of \(\{\ket{M_{a_1}},\ket{M_{a_2}}, \dots, \ket{M_{a_n}}\}\). Thus,  after the measurement interaction, the measurer being in the state \(\ket{M_{a_i}}\) has the physical meaning `the observable \(\hat{A}\) has value \(a_i\)'. Notably, Eq.~\eqref{eq:measurement} is a unitary process and replaces the collapse postulate.\footnote{The measurement interaction defined in Eq.~\eqref{eq:measurement} is unitary on the domain spanned by \( \{ \ket{M_0} \ket{a_i} : 1 \leq i \leq n \}\). Because the measurement interaction is unitary on this subspace, it can be extended to be unitary on the entire Hilbert space, as shown in Exercise 2.67 of Ref.~\cite{nielsen2010quantum}.}

In cases where the system being measured is in a superposition of eigenstates of \(\hat{A}\), the measurement protocol still performs the desired measurement, but there is then no definite measurement outcome, as several outcomes will be observed by the measurer simultaneously. Consider, for instance, the following measurement of a system in superposition, which occurs between times \(t\) and \(t+1\), 
\begin{equation}
   \ket{\Psi(t)}=\sum_{i=1}^n\alpha_i \ket{M_0} \ket{a_i} \overset{\text{measurement}}{\xlongrightarrow{\hspace{1cm}}}  \ket{\Psi(t+1)}=\sum_{i=1}^n \alpha_i \ket{M_{a_i}} \ket{a_i},
\end{equation}
where, for simplicity, \(\alpha_1,...,\alpha_n\) are positve real coefficients such that the initial and final states of the composite system are normalised. Evidently, the state vector \(\ket{\Psi(t)}\) is a product state, and due to the measurement, both the measurer and the systems being measured become entangled, ending up in the entangled state \(\ket{\Psi(t+1)}\). During this measurement, more than one outcome was recorded by the measurer because each of the \(n\) states in \(\{\ket{M_{a_1}},\ket{M_{a_2}}, \dots, \ket{M_{a_n}}\}\) appear in \(\ket{\Psi(t+1)}\), and each of these states corresponds to the measurer having recorded a distinct result. The existence of these different instances of the measurer is why Everett's construction is also called the \textit{many-worlds interpretation} of quantum theory and why the whole of physical reality, as described by quantum theory, is referred to as the \textit{multiverse}.\footnote{The name `many worlds' can be slightly misleading, as Everett's construction does not assume there are other worlds or branches. Instead, all that one has to assume is that quantum systems invariably evolve unitarily, with the other branches of the state vector arising as an emergent consequence of this more fundamental assumption.} 
 
 Of particular importance to local realism is that the measurement interaction is a unitary process so that, at the level of the state vector, there is no hint of collapse and the non-locality that it entails. However, Everett's construction is here formulated in the Schrödinger picture, and this picture remains problematic because it is not a separable description. This non-separability affects Everett's \textit{relative states}; for instance, the state of the composite system \( \ket{\Psi(t+1)} \) relative to the observable \(\hat{A}\) having value \(a_i\) is defined as
\begin{equation}
  \frac{  \ket{a_i} \langle a_i |  \Psi(t+1) \rangle }{ | \langle a_i |  \Psi(t+1) \rangle|} = \ket{M_{a_i}} \ket{a_i}.
\end{equation}
It is in the relative states that systems \textit{appear} to have collapsed since the relative states mimic the state vector collapse expressed in \eqref{eq:collapse}, yet the universal state vector \(\ket{\Psi(t+1)}\) has evidently not undergone collapse. Thus, the two different types of dynamical processes – namely, unitary evolution and collapse – have been reduced solely to unitary evolution. 

What remains problematic for local realism is that the relative state of one system can be affected by a measurement performed on another distant system. Another way of putting this is that the relative states necessarily represent everything in existence relative to some observable having a certain value, and this makes it seem as if there is action at a distance \cite{Vaidman2021}. What is desired for a local description is that entangled quantum systems can be described as local `bubbles' so that the goings-on in the interior of these bubbles do not affect the description of anything in their exterior. In Sec.~\ref{sec:relativedescriptors}, I will provide such a description.

\section{The mathematics of local realism} \label{sec:locality}
The informal account of local realism provided in Sec.~\ref{introduction} has been formalised in an axiomatic framework by Raymond-Robichaud, whose work (see Refs.~\cite{raymond2021local, raymond2017equivalence}) will be reviewed in this section.\footnote{Readers seeking the full technical details are encouraged to consult the original works.} In this axiomatisation, henceforth referred to simply as \textit{local realism}, the local description of a system is given by its \textit{noumenal states}, defined as representations that satisfy both separability and no action at a distance. The term `noumenal state' is borrowed from Kantian philosophy, where it refers to the realm of things as they are in themselves. It contrasts with the phenomenal world, the realm of appearances mediated through our senses.

As such, any system \(A\) is assumed to have a nonempty space of noumenal states \(\textsf{Noumenal-Space}^{A}\). The noumenal state of a composite system \(AB\) can be separated into the noumenal states of its subsystems, \(A\) and \(B\). For instance, consider the noumenal state \( N^{AB} \in \textsf{Noumenal-Space}^{AB} \) of the composite system \(AB\); then separability requires that the noumenal state \(N^{AB}\) admits a decomposition into states \(N^A\) and \(N^B\), and that it must be possible to recombine \(N^A\) and \(N^B\) into \(N^{AB}\) again. This can be schematically summarised as
\begin{equation}
N^{AB} \xrightarrow{\text{decompose}}  ( N^{A},N^{B}) \xrightarrow{\text{recombine}} N^{AB}.
\end{equation}
The functions that act on the noumenal state \(N^{AB}\) to decompose it into \(N^A\) and \(N^B\) are termed the \textit{noumenal projectors}. In particular, for any system \(A\) that is a subsystem of \(AB\), there exists a surjective function 
\begin{align}
\pi_A:  \textsf{Noumenal-Space}^{AB} \rightarrow \textsf{Noumenal-Space}^{A}
\end{align}
whose effects on \(N^{AB}\) is that
\begin{align}
\pi_A(N^{AB}) & = N^A .
\end{align}
That is to say that, given the noumenal state of the composite system \( N^{AB} \), the noumenal projectors produce the states \( N^{A}\) of subsystem \(A\). Two states \( N^{A}\) and \(N^{B}\) are said to be \textit{compatible} if there exists \(N^{AB}\) such that \(N^A = \pi_A(N^{AB})\) and \(N^B = \pi_B(N^{AB})\).

The operation that recombines the noumenal states is an injective function \(\odot : \textsf{Noumenal-Space}^{A} \times  \textsf{Noumenal-Space}^{B}  \rightarrow \textsf{Noumenal-Space}^{AB} \), called the \textit{noumenal product}. When the noumenal states \(N^A\) and \(N^B\) are compatible, the effect of the noumenal product on \(N^A\) and \(N^B\) is that
\begin{align} 
N^A \odot N^B = N^{AB}.
\end{align}
The combined effect of the noumenal projectors and the noumenal product is captured by the following relation:
\begin{equation} \label{eq:separability}
\pi_{A}(N^{AB}) \odot \pi_{B}(N^{AB})  = N^{AB}.
\end{equation}

The noumenal state of an arbitrary system, \(A\), evolves according to certain dynamical laws, which are implemented by operations that act on the noumenal states. As such, each system is assumed to have a group of operations \(\textsf{Operations}^A\) together with some action \(\star_A : \textsf{Operations} \times \textsf{Noumenal-Space}^A \to \textsf{Noumenal-Space}^A\) such that the effect of \(U \in \textsf{Operations}^A\) on \(N^{A}\) will produce a new noumenal state, denoted as \(U \star N^{A}\).  Local realism imposes constraints on how the evolution of two disjoint subsystems \(A\) and \(B\) affects the noumenal state of a composite system \(AB\). In particular, a \textit{product of operations}, denoted \(\times\), must exist, which combines operations on the subsystems into operations on the composite system. Consider, for instance, operations \(V \in \textsf{Operations}^A\) and \(W \in \textsf{Operations}^B \); then there is assumed to exists a product of operations \(V \times W \in \textsf{Operations}^{AB}\), whose effect on any noumenal state \(N^{AB} = N^A \odot N^B\) is that
\begin{align} \label{eq:localevolution}
(V \times W) \star (N^A \odot N^B )  = (V \star N^A ) \odot (W \star N^B ).
\end{align}
This rigorously captures Einstein's less formal definition of there being no action at a distance because we see that the operation \(W\) that is performed on \(B\) leaves the noumenal state of \(A\) unaffected, and \(V\) leaves the state of \(B\) unaffected, just as Einstein required. Hence, the noumenal states \( N^A\) and \(N^B\) correspond to the local description of reality that Einstein sought and desired. 

The noumenal state of a system \(A\) provides a complete description of that system at a particular instant, so it should allow one to deduce any locally observable property of that system at that instant. 
The locally observable properties of a system are determined by its \textit{phenomenal state}, so each system \(A\) is assumed to have a nonempty set \(\textsf{Phenomenal-Space}^A\), together with a structure-preserving surjective function (i.e., an epimorphism):
\begin{align}
\varphi : \textsf{Noumenal-Space}^A \to \textsf{Phenomenal-Space}^A,
\end{align}
ensuring that for each phenomenal state of a system, there exists an \(N^A\) such that \(\varphi(N^A)\) is that phenomenal state. Crucially, the phenomenal states of a system do not have to satisfy separability, as separability is a distinct property of the noumenal states.

I define \textit{the principle of locality} as the assumption that a theory of physics should satisfy local realism – i.e., that a physical theory should have noumenal states that adhere to the constraints of Eqs.~\eqref{eq:localevolution} and \eqref{eq:separability}. Notably, a quantum system may be described by non-local models as well as local ones; the principle only demands that at least one such local model exists.

Neither the Schrödinger picture nor the density matrix formalism is a local description since neither is separable. For example, if \( \hat{\rho}_{AB}\) is the density matrix of a two-qubit composite system, then although \( \hat{\rho}_{AB}\) can be decomposed via the trace into two density matrices \( \hat{\rho}_A\) and \(\hat{\rho}_B\) describing the respective qubits, those reduced density matrices cannot generally be recombined back into \( \hat{\rho}_{AB}\). A particular example of a proposed noumenal product that fails to satisfy Eq.~\eqref{eq:separability} for all possible density matrices is the tensor product, since when the two qubits are entangled, 
\begin{equation}
\hat{\rho}_A \otimes \hat{\rho}_B  \not= \hat{\rho}_{AB}.
\end{equation}
More generally, besides the tensor product, there is no function that allows one to recover an arbitrary density matrix \( \hat{\rho}_{AB} \) from its reduced density matrices \( \hat{\rho}_A \) and \( \hat{\rho}_B \). Consequently, the density matrix formalism does not satisfy the requirements of local realism~\cite{bedard2021abc}. A similar line of reasoning can be used to show that the Schrödinger picture state vector does not provide a separable description (for such a proof, see Ref.~\cite{raymond2021local}), implying that the Schrödinger and Heisenberg pictures constitute fundamentally different descriptions -- as argued by Bédard~\cite{Bedard2025} in this volume.

\section{The Heisenberg picture}
Although neither the Schrödinger picture nor the density matrix formalism is a local description, quantum theory \textit{does} satisfy the principle of locality because at least one local description of quantum systems exists – local theories often admit non-local descriptions besides a local one (see Appendix~\ref{sec:derivation}).

That local description of quantum theory is the Heisenberg picture. I will elucidate this picture's foundations by using it to describe a network of \(n\) qubits. There are two main reasons for the choice to study a network of qubits: the first is historical, as much research on locality has taken place within the context of quantum information theory – see, for instance, Refs.\cite{deutsch2000information, bedard2021cost, kuypers2021everettian, raymond2021local}. Secondly, a network of qubits is a simple system in that it has a finite-dimensional Hilbert space, making it more amenable to study. 

As such, consider a quantum computational network,  \(\mathcal{N}\), consisting of \(n\) qubits, denoted \(\mathcal{Q}_1, \dots, \mathcal{Q}_n \). In the Heisenberg picture, the state vector is stationary, and instead, it is the observables of the system that are time-dependent. In particular, a qubit \(\mathcal{Q}_a\), at some time \(t\), is represented by a triple of observables that are called its \textit{descriptors}
\begin{equation*}
    \hat{\boldsymbol{q}}_a(t) \stackrel{\text{def}}{=} (\hat{q}_{ax}(t),\hat{q}_{ay}(t),\hat{q}_{az}(t)),
\end{equation*}
which adhere to the Pauli algebra at any time \(t\), meaning that they are Hermitian and satisfy the following algebraic relation:
\begin{equation} \label{pauli}
    \hat{q}_{ai}(t)\hat{q}_{aj}(t) = \hat{1} \delta_{ij} + i \epsilon_{ij}^{\ \ k}\hat{q}_{ak} (t) \qquad \qquad (\forall i,j \in \{x,y,z\}).
\end{equation} Notably, I rely on Einstein's summation convention so that the otherwise undefined index \(k\), which appears both as a superscript and a subscript in a product, is summed over the complete set of its possible values. For the index \(k\), this set is \(\{x,y,z\}\). 
Additionally, \(\hat{1}\) is the unit observable, which is defined so that its product with all other observables leaves the latter unaffected, i.e.
\begin{equation} \label{identity}
\qquad  \hat{1}\hat{q}_{aj}(t) =\hat{q}_{aj}(t)\hat{1} = \hat{q}_{aj}(t) \qquad \qquad  (\forall j \in \{x,y,z\}).
\end{equation}

The Pauli algebra is not commutative since, for instance, \( \hat{q}_{ax}(t)\hat{q}_{ay}(t) \neq \hat{q}_{ay}(t)\hat{q}_{ax}(t)\). This makes the descriptors \textit{q-numbers} in Dirac’s terminology \cite{dirac1981principles}, meaning they are objects that can be added, with addition being associative and commutative, and multiplied, with multiplication being associative but not necessarily commutative. In contrast, conventional numbers, such as real and complex numbers, always commute under multiplication and are called \textit{c-numbers}.

The \(q\)-numbers \(\hat{q}_{ax}(t), \hat{q}_{ay}(t), \hat{q}_{az}(t)\) can be combined by addition, multiplication, and complex scaling to form new \(q\)-numbers. Among these are the observables, which are Hermitian combinations of these three q-numbers and the identity. The most general observable constructed from \(\hat{q}_{ax}(t)\), \(\hat{q}_{ay}(t),\) and \( \hat{q}_{az}(t)\) by addition, multiplication, and complex scaling has the form
\begin{equation} \label{eq:observable}
    \hat{A}(t) \stackrel{\text{def}}{=}  a_0 \hat{1} + a_x\hat{q}_{ax}(t)+a_y\hat{q}_{ay}(t)+a_z\hat{q}_{az}(t),
\end{equation}
where the coefficients \(a_0,a_x,a_y,\) and \(a_z\) are real numbers. This form is exhaustive because the Pauli algebra ensures that any product of \(\hat{q}_{ax}(t),\hat{q}_{ay}(t),\hat{q}_{az}(t)\) reduces to a linear combination of \( \hat{q}_{ax}(t),\hat{q}_{ay}(t), \hat{q}_{az}(t)\), and \(\hat{1}\). 

Since any observable of the qubit can be expressed in terms of the components of \( \hat{\boldsymbol{q}}_a(t) \), as shown in Eq.~\eqref{eq:observable}, one only has to keep track of this triple in order to keep track of any other observable of \(\mathcal{Q}_a\). In fact, even \( \hat{\boldsymbol{q}}_a(t) \) contains superfluous information since, due to the Pauli algebra, one of its components can be generated via a product of the remaining two, which is why some authors opt to keep track only of a pair of descriptors \cite{bedard2021abc}. For the sake of clarity, I will always keep track of all three of the components of \( \hat{\boldsymbol{q}}_a(t) \).

\subsection{Expectation values and phenomenal states}
A complete description of a qubit, \( \mathcal{Q}_a\), is provided by its triple of descriptors \( \hat{\boldsymbol{q}}_a(t) \) together with their expectation values. Those expectation values are determined by the \textit{Heisenberg state}, denoted \(\hat{\varrho} \). The Heisenberg state is a constant q-number, meaning it remains unchanged even as the qubit evolves. Specifically, \( \hat{\varrho}\)  is defined as the dyadic of the initial state vector of the network: if at time \(t=0\), the state of the network in the Schrödinger picture is \(\ket{\psi}\), then the Heisenberg state is given by \(\hat{\varrho} \stackrel{\text{def}}{=} \ket{\psi}\bra{\psi}\). The role of the Heisenberg state is to determine the expectation values of the qubit's observables, as follows:
\begin{equation}
    \langle \hat{q}_{aj} (t) \rangle_{\hat{\varrho}} \stackrel{\text{def}}{=} \text{Tr}_{\mathcal{N}} (\hat{q}_{aj} (t) \hat{\varrho})  \qquad \qquad (j \in \{x,y,z\}),
\end{equation}
where \(\text{Tr}_{\mathcal{N}}\) is the trace over the \(2^n\)-dimensional Hilbert space \(\mathcal{H}_{\mathcal{N}}\) of the \(n\)-qubit network, \(\mathcal{N}\).\footnote{The Hilbert space of the  \(n\)-qubit network is the tensor product of the 2-dimensional Hilbert spaces of the individual qubits.} 

Notably, the expectation value of a sum of operators is the sum of their expectation values. 
Since any observable of \(\mathcal{Q}_a\) is, by Eq.~\eqref{eq:observable}, a linear combination of 
\( \hat{q}_{ax}(t), \hat{q}_{ay}(t), \hat{q}_{az}(t) \), and \(\hat{1}\), 
its expectation value is determined entirely by the expectation values of these four q-numbers. 
Thus, keeping track of these four expectation values suffices to give a complete account of what is locally observable about \(\mathcal{Q}_a\), i.e. its phenomenal state. 
In particular, the phenomenal state is
\begin{align}
\langle \hat{\boldsymbol{q}}_a(t) \rangle 
  &= \big( \langle \hat{q}_{ax}(t) \rangle,\;
            \langle \hat{q}_{ay}(t) \rangle,\;
            \langle \hat{q}_{az}(t) \rangle \big),
\end{align}
where we do not need to keep track of \(\hat{1}\)'s expectation value explicitly, since it is always \(1\).

The expectation value alone does not inform us, at least directly, how much a measurement result of \(\hat{q}_{az}(t)\) is likely to deviate from \(\langle \hat{q}_{az}(t) \rangle_\varrho \). This spread in the results is represented by the \textit{variance}; the variance of an arbitrary observable \(\hat{O}(t)\) is defined as 
\begin{align}
    V_{\hat{\varrho}}(\hat{O} (t))  \stackrel{\text{def}}{=}  \langle \hat{O}(t)^2 \rangle_{\hat{\varrho}}  - \langle  \hat{O}(t)\rangle_{\hat{\varrho}}^2.
\end{align}
The variance is a non-negative real number, and the greater the value of the variance, the greater the spread in the measurement results of \(\hat{O} (t)\). Conversely, the smaller the variance, the smaller the spread. In the ideal case in which the variance is zero, i.e. \(V_{\hat{\varrho}}(\hat{O} (t)) =0\), the observable \(\hat{O} (t)\) has a definite value equal to \(\langle \hat{O} (t) \rangle_{\hat{\varrho}}\), in which case the observable is said to be \textit{sharp} with that value.

\subsection{Quantum theory's noumenal and phenomenal states}\label{sec:noumenal}
Thus far, I have discussed the algebraic relations of the descriptors of an individual qubit, as well as the expectation values of such descriptors. However, as it stands, the algebraic relations between the descriptors of two distinct qubits, such as \(\hat{q}_{ai} (t)\) and \(\hat{q}_{bj} (t)\), for \(i,j \in \{x,y,z\}\), are not yet specified. This is why the additional assumption is required that at any time \(t\), for any pair of qubits \(\mathcal{Q}_a\) and \(\mathcal{Q}_b\) such that \(a\not= b\), their descriptors are assumed to commute
\begin{equation}  \label{eq:commutator}
    [ \hat{q}_{ai}(t), \hat{q}_{bj}(t)] =0 \qquad \qquad (\forall i,j \in \{x,y,z\} ).
\end{equation}
This will be referred to as the \textit{commutation constraint}. Jointly, the commutation constraint (as given in Eq.~\eqref{eq:commutator}) and the Pauli algebra (as given in Eq.~\eqref{pauli}) fully specify the algebraic relations of the descriptors of any number of qubits.

The noumenal state of an arbitrary qubit \(\mathcal{Q}_a\) is the pair \((\hat{\boldsymbol{q}}_a(t), \hat{\varrho})\). It provides a complete account of the qubit, including its phenomenal state, because the triple of expectation values \(\langle \hat{\boldsymbol{q}}_a(t) \rangle\) can be computed from this pair. Thus, in the \(n\)-qubit network \(\mathcal{N}\), the noumenal states of each of the \(n\) qubits separately are \((\hat{\boldsymbol{q}}_1(t), \hat{\varrho}), (\hat{\boldsymbol{q}}_2(t), \hat{\varrho}),...,(\hat{\boldsymbol{q}}_n(t), \hat{\varrho}) \), where the descriptors of any two of these noumenal states satisfy Eq.~\eqref{eq:commutator} and each triple of descriptors separately obeys the Pauli algebra, as expressed in Eq.~\eqref{pauli}. These \(n\) noumenal states jointly describe the state of the whole network. For instance, to compute the expectation value of a joint observable, such as the product \(\hat{q}_{aj} (t) \hat{q}_{bj}  (t)\), one merely has to use the descriptors of the individual qubits and the Heisenberg state:
\begin{equation} \label{eq:expectation.joint}
 \langle  \hat{q}_{aj} (t) \hat{q}_{bj}  (t)\rangle_{\hat{\varrho}}=   \text{Tr}_{\mathcal{N}}( \hat{q}_{aj} (t) \hat{q}_{bj}  (t)\hat{\varrho} ).
\end{equation}
A general observable of the network at time \(t\) is a polynomial with real-valued coefficients in the \(3n\) descriptors \(\hat{q}_{1x}(t),\hat{q}_{1y}(t),\hat{q}_{1z}(t), \dots,\hat{q}_{nx}(t),\hat{q}_{ny}(t),\hat{q}_{nz}(t) \). Clearly, one can generate any such polynomial from the \(n\) triples \(\hat{\boldsymbol{q}}_1(t),\dots, \hat{\boldsymbol{q}}_n(t)\), just as one could generate the complete set of single-qubit observables of \(\mathcal{Q}_a\) from \(\hat{\boldsymbol{q}}_a(t)\) alone. The expectation value of any such observable of the network is determined by that observable together with the network's Heisenberg state \(\hat{\varrho}\), as for instance shown in Eq.~\eqref{eq:expectation.joint}. As such, the noumenal state of the network of \(n\) qubits is fully described by the \((n+1)\)-tuple  \((\hat{\boldsymbol{q}}_1(t),\hat{\boldsymbol{q}}_2(t),\dots, \hat{\boldsymbol{q}}_n(t), \hat{\varrho})\).

To demonstrate that these proposed noumenal states satisfy separability, one can define the noumenal product of the noumenal states of two qubits, \(\mathcal{Q}_a\) and \(\mathcal{Q}_b\), as follows:
\begin{equation}
(\hat{\boldsymbol{q}}_a(t), \hat{\varrho})  \odot (\hat{\boldsymbol{q}}_b(t), \hat{\varrho}) \stackrel{\text{def}}{=} (\hat{\boldsymbol{q}}_a(t), \hat{\boldsymbol{q}}_b(t), \hat{\varrho}),
\end{equation}
where \((\hat{\boldsymbol{q}}_a(t), \hat{\boldsymbol{q}}_b(t), \hat{\varrho})\) is the noumenal state of a composite system consisting of the two qubits \(\mathcal{Q}_a\) and \(\mathcal{Q}_b\). In like manner, the noumenal state of the composite system can be decomposed into that of \(\mathcal{Q}_a\) and \(\mathcal{Q}_b\) via the following noumenal projectors:
\begin{align}
\pi_a \bigl( (\hat{\boldsymbol{q}}_a(t), \hat{\boldsymbol{q}}_b(t), \hat{\varrho})  \bigr) & \stackrel{\text{def}}{=}  (\hat{\boldsymbol{q}}_a(t), \hat{\varrho}), \\ 
\pi_b \bigl( (\hat{\boldsymbol{q}}_a(t), \hat{\boldsymbol{q}}_b(t), \hat{\varrho}) \bigr) & \stackrel{\text{def}}{=} (\hat{\boldsymbol{q}}_b(t), \hat{\varrho}).
\end{align}
This straightforwardly generalises to systems of more than two qubits, implying that the proposed noumenal states of the qubits satisfy separability. Hence, the Heisenberg picture is a separable description of quantum theory, as promised. This connection between Raymond-Robichaud’s work and that of Deutsch \& Hayden is discussed in \cite{bedard2021cost}, albeit by a different method than the one employed here.

Since the noumenal state of each qubit contains a copy of the network's Heisenberg state \(\hat{\varrho}\), some authors have expressed concerns about the possible non-locality of the Heisenberg picture \cite{timpson2005nonlocality}. The concern is that the Heisenberg state describes the network as a whole and, therefore, shouldn't be included in the noumenal state of each qubit individually. However, the Heisenberg state is a constant in unitary quantum theory, so its presence within the noumenal states of each qubit does not pose a problem for local realism. As Deutsch \cite{deutsch2024private} summarises this issue, `to worry about the Heisenberg state describing distant systems is exactly like worrying about the number \(\pi\) describing distant circles.' That is to say that both \(\hat{\varrho}\) and \(\pi\) are constants, meaning that operations performed on spacelike separated systems do not affect them; it is unproblematic to include such global entities in the noumenal states of individual qubits.

\subsection{The dynamics of descriptors} \label{dynamics}
Not only does the Heisenberg picture satisfy separability, but its dynamics also ensure there is no action at a distance, meaning that Eq.~\eqref{eq:localevolution} is satisfied. I have not yet discussed dynamics since, in the preceding section, only static systems were considered, namely qubits at a particular instant \(t\).

In the network model, qubits evolve due to the gates that are enacted on them. These gates take as input the qubits' noumenal states at time \(t\) and output their noumenal states at a later time, \(t+1\); the fact that the gate acts during one unit of time is a simplifying idealisation, but we lose no generality by assuming it. A general gate \(G\) is implemented at time \(t\) by a unitary transformation \(U_G\), where by virtue of being unitary, \(U_G\) must adhere to the constraint that 
\begin{equation} \label{eq:identity}
    U^{\dagger}_G U_G = U_G U_G^{\dagger} = \hat{1}.
\end{equation}
In the network model, the unitary \(U_G\) at time \(t\) is a complex polynomial in the \(3n\) descriptors \(\hat{q}_{1x}(t),\hat{q}_{1y}(t),\hat{q}_{1z}(t), \dots,\hat{q}_{nx}(t),\hat{q}_{ny}(t),\hat{q}_{nz}(t) \), so I will say that the unitary is a function of the descriptors, and to accentuate that the unitary is a function of the descriptors, I shall denote the unitary at time \(t\) as \( U_G(\hat{\boldsymbol{q}}_1(t),\dots, \hat{\boldsymbol{q}}_n(t))\). Moreover, if the unitary depends on only a subset of the descriptors, I will include only those specific descriptors in its argument. Thus, the evolution of a qubit's descriptors between times \(t\) and \(t+1\) due to the enactment of the general gate \(G\) is
\begin{equation} \label{eq:unitaries}
    \hat{\boldsymbol{q}}_{a}(t+1) = U_G^{\dagger} (\hat{\boldsymbol{q}}_1(t),\dots, \hat{\boldsymbol{q}}_n(t))  \hat{\boldsymbol{q}}_{a}(t) U_G(\hat{\boldsymbol{q}}_1(t),\dots, \hat{\boldsymbol{q}}_n(t)).
\end{equation}
This expression is to be understood to mean that the unitary \(U_G\) conjugates each component of the triple \(\hat{\boldsymbol{q}}_{a}(t)\) individually.

Unitary evolution has the important property of preserving the algebraic relationships of the descriptors. For instance, due to unitary evolution, the Pauli algebra is a constant of the motion, so the descriptors adhere to that algebra at any time \(t\). To prove this, assume that \( \hat{\boldsymbol{q}}_{a}(t)\) adheres to the Pauli algebra at time \(t\). If \( \hat{\boldsymbol{q}}_{a}(t)\) evolves unitarily between \(t\) and \(t+1\), then the triple will also adhere to that same algebra at time \(t+1\) since
\begin{align} \label{eq:evolution}
    \hat{q}_{ai}(t+1) \hat{q}_{aj}(t+1) & = U_G^{\dagger} \hat{q}_{ai}(t) U_G U_G^{\dagger} \hat{q}_{aj}(t) U_G, \\
& = U_G^{\dagger} \hat{q}_{ai}(t) \hat{q}_{aj}(t) U_G, \\
& = U_G^{\dagger} ( \hat{1} \delta_{ij} + i \epsilon_{ij}^{\ \ k}\hat{q}_{ak} (t) ) U_G, \\
& = \hat{1} \delta_{ij} + i \epsilon_{ij}^{\ \ k}\hat{q}_{ak} (t+1),
\end{align}
where I have dropped \(U_G\)'s dependence on the descriptors for clarity. Any other algebraic relation of the descriptors is also invariant. For instance, descriptors that mutually commute at time \(t\) will still do so at time \(t+1\), as can be readily verified by a similar proof as that used above, implying that under unitary evolution, the algebra of the system is a constant of the motion.

We have already seen how the proposed noumenal states of \(\mathcal{N}\) satisfy separability, which is one of two key requirements for local realism (see Sec.~\ref{sec:locality}). The second requirement for local realism is that there is no action at a distance: if a gate is enacted on one qubit and not on another, the noumenal state of the unaffected qubit should be unchanged. The proposed noumenal states adhere to this requirement, too. Consider, for example, a general single qubit gate enacted on \(\mathcal{Q}_b\). The unitary for such a gate is
\begin{equation} \label{eq:single-qubit-unitary}
    U_{b} ( \boldsymbol{\hat{q}}_b(t) ) = a_0 \hat{1} + a_x\hat{q}_{ax}(t)+a_y\hat{q}_{ay}(t)+a_z\hat{q}_{az}(t),
\end{equation}
where the coefficients \(a_0,a_x,a_y,a_z\) are complex numbers such that \(U_{b}\) is a unitary. Henceforth, the unitary for a general single-qubit gate acting on \(\mathcal{Q}_b\) will be denoted by \(U_b\), where its dependence on \(\boldsymbol{\hat{q}}_b(t)\), as shown in Eq.~\eqref{eq:single-qubit-unitary}, will be assumed implicitly. Due to the commutation constraint of Eq.~\eqref{eq:commutator}, this unitary commutes with the descriptors of all other qubits. Consequently, the effect of this gate on another qubit, \(\mathcal{Q}_a\), where \(a\not=b\), is
\begin{equation} \label{eq:unaffected}
    \hat{\boldsymbol{q}}_{a}(t+1) = U^{\dagger}_{b} \hat{\boldsymbol{q}}_{a}(t) U_{b} = \hat{\boldsymbol{q}}_a(t).
\end{equation}
Here I used Eqs.~\eqref{eq:identity}, \eqref{eq:commutator} and \eqref{eq:single-qubit-unitary}.
This fact can be used to construct a formal proof that the no-action-at-a-distance requirement of Eq.~\eqref{eq:localevolution} is satisfied. For this proof, one should define a product of operations for the qubits' noumenal states. Consider applying a single-qubit gate \(U_a\) to \(\mathcal{Q}_a\) and another single-qubit gate \(U_b\) to \(\mathcal{Q}_b\). The effects of these gates on the noumenal states of the respective qubits are defined as:
\begin{align} \label{eq:single.qubit.operations}
    U_a \star ( \boldsymbol{\hat{q}}_a(t), \hat{\varrho}) &  \stackrel{\text{def}}{=} ( U^\dagger_a \boldsymbol{\hat{q}}_a(t) U_a, \hat{\varrho}), \\
 U_b \star ( \boldsymbol{\hat{q}}_b(t), \hat{\varrho}) &  \stackrel{\text{def}}{=} ( U^\dagger_b \boldsymbol{\hat{q}}_b(t) U_b, \hat{\varrho}).
\end{align}
These are single-qubit gates performed separately on each qubit. As such, \(U_a\) is a function solely of \(\boldsymbol{\hat{q}}_a(t)\), and \(U_b\) is a function solely of \(\boldsymbol{\hat{q}}_b(t)\). A two-qubit gate, \(U_{ab}\), that acts on \(\mathcal{Q}_a\) and \(\mathcal{Q}_b\) has an effect on the noumenal state of this two-qubit system that can similarly be defined as 
\begin{align} \label{eq:two.qubits.operations}
    U_{ab} \star ( \boldsymbol{\hat{q}}_a(t), \boldsymbol{\hat{q}}_b(t), \hat{\varrho}) &  \stackrel{\text{def}}{=} ( U^\dagger_{ab} \boldsymbol{\hat{q}}_a(t) U_{ab}, U^\dagger_{ab} \boldsymbol{\hat{q}}_b(t) U_{ab}, \hat{\varrho}).
\end{align}
Note that in general, the two-qubit gate \( U_{ab}\) is a function of both \( \boldsymbol{\hat{q}}_a(t) \) and \(\boldsymbol{\hat{q}}_b(t) \). The desired product of operations is then simply the product of the individual unitaries
\begin{align} \label{eq:product.of.operations}
    ( U_a \times U_b ) \star  ( \boldsymbol{\hat{q}}_a(t), \boldsymbol{\hat{q}}_b(t), \hat{\varrho}) & \stackrel{\text{def}}{=}  ( U_a U_b ) \star  ( \boldsymbol{\hat{q}}_a(t), \boldsymbol{\hat{q}}_b(t), \hat{\varrho}), \\
&  = ( U^\dagger_b U^\dagger_a  \boldsymbol{\hat{q}}_a(t) U_a U_b,  U^\dagger_b U^\dagger_a \boldsymbol{\hat{q}}_b(t) U_a U_b, \hat{\varrho}), \\
&  = ( U^\dagger_a  \boldsymbol{\hat{q}}_a(t) U_a, U^\dagger_b \boldsymbol{\hat{q}}_b(t) U_b, \hat{\varrho}), \\
& =  \left ( U_a \star ( \boldsymbol{\hat{q}}_a(t), \hat{\varrho}) \right ) \odot \left ( U_b \star ( \boldsymbol{\hat{q}}_b(t), \hat{\varrho}) \right )
\end{align}
Here, I have used the fact that \(U_a\) is a function of \(\boldsymbol{\hat{q}}_a(t)\) so that this unitary commutes with \(\boldsymbol{\hat{q}}_b(t)\), due to the commutation constraint of Eq.~\eqref{eq:commutator}; for the same reason, \(U_b\) commutes with \(\boldsymbol{\hat{q}}_a(t)\), and \(U_a\) and \(U_b\) also commute with each other. Note also that \(U_a U_b \) is simply the product of the two unitaries \(U_a\) and \(U_b\). The existence of a product of operations shows that the proposed noumenal states comply with Eq.~\eqref{eq:localevolution}, completing the proof that quantum theory conforms to local realism.

\section{Relative descriptors} \label{sec:relativedescriptors}
It is crucial for local realism that the dynamics of quantum systems are exclusively unitary. For instance, in the Heisenberg picture, collapse is typically modelled as a change in the Heisenberg state, and since the noumenal state \( (\hat{\boldsymbol{q}}_{a}(t), \hat{\varrho})\) of an arbitrary qubit \(\mathcal{Q}_a\) contains a copy of the Heisenberg state of the network, this change in the Heisenberg state due to a measurement of one qubit must affect the noumenal state of any other qubit, too. This would imply action at a distance. Everettian quantum theory, by offering a fully unitary account, therefore plays a key role in local realism. However, as of yet, it is unclear how Everett’s relative states are to be defined in the Heisenberg picture, and without such a description, it is challenging to account for measurement in a local way. How can one provide a local account of Everett’s relative state construction? And how can one explain the non-locality in the EPR argument, which arises from the apparently non-local effects of measurement?

To that end, consider a model measurement in which one qubit measures another qubit. The measurement interaction is modelled by the CNOT gate. The target qubit of the CNOT acts as a measurer, and the control qubit is the system being measured. Let us call the control qubit \(\mathcal{Q}_a\) and the target \(\mathcal{Q}_b\). At a time \(t\), the qubits have the following descriptors,
\begin{align}
    \hat{\boldsymbol{q}}_a (t) & = (\hat{q}_{ax}(t),\hat{q}_{ay}(t),\hat{q}_{az}(t)), \\
\hat{\boldsymbol{q}}_b (t) & = (\hat{q}_{bx}(t),\hat{q}_{by}(t),\hat{q}_{bz}(t)).
\end{align}
When the CNOT acts between times \(t\) and \(t+1\), it has the following characteristic effect on \(\mathcal{Q}_a\) and \(\mathcal{Q}_b\):
\begin{align} \label{eq:CNOT}
\left\{
\begin{aligned}
    \hat{\boldsymbol{q}}_a (t)\\
\hat{\boldsymbol{q}}_b (t)
\end{aligned} \right \}
\overset{\text{CNOT}}{\longrightarrow} \left\{
\begin{aligned}
    \hat{\boldsymbol{q}}_a (t+1) & = (\hat{q}_{ax}(t),\hat{q}_{ay}(t)\hat{q}_{bz}(t),\hat{q}_{az}(t)\hat{q}_{bz}(t)) \\
\hat{\boldsymbol{q}}_b (t+1) & = (\hat{q}_{bx}(t)\hat{q}_{ax}(t),\hat{q}_{by}(t)\hat{q}_{ax}(t),\hat{q}_{bz}(t))
\end{aligned} \right \}
\end{align}

When the \(z\)-observable of the control qubit is sharp with value \(\phi \in \{1,-1\}\), and the target qubit has a \(z\)-observable that is sharp with the receptive value \(1\), then the CNOT gate copies the value \(\phi\) of the control's \(z\)-observable onto that of the target qubit, as is evident from the effect of the CNOT on the qubits' expectation values. To make the notation more transparent later on, I shall assume that the measurement interaction starts at \(t_0-1\) so that it ends at time \(t_0\). The effect of the CNOT for this proposed measurement between \(t_0-1\) and \(t_0\) is 
\begin{align}
\left\{
\begin{aligned}
\langle \hat{q}_a(t_0-1) \rangle_{\hat{\varrho}} &= (0, 0, 1) \\
\langle \hat{q}_b(t_0-1) \rangle_{\hat{\varrho}} &= (0, 0, \phi)
\end{aligned}
\right\}
\overset{\text{CNOT}}{\longrightarrow}
\left\{
\begin{aligned}
\langle \hat{q}_a(t_0) \rangle_{\hat{\varrho}} & = (0, 0, \phi) \\
\langle \hat{q}_b(t_0) \rangle_{\hat{\varrho}} & = (0, 0, \phi)
\end{aligned}
\right\}
\end{align}
Evidently, the bit of information, \(\phi\), that is initially stored in the control qubit is copied onto the target, so at time \(t_0\), the values \(1\) and \(-1\) of the target qubit's \(z\)-observable have the physical meaning `it was a \(1\)' and `it was a \(-1\)', respectively, which is why the CNOT gate models a measurement interaction.

When the control qubit has a \(z\)-observable that is non-sharp (meaning its variance is non-zero), the CNOT still performs a measurement, but the measurement is such that the measurer will record multiple results simultaneously. Consider, for example, the case in which the control qubit initially (i.e. at \(t_0-1\), the instant before the measurement) has a non-sharp \(z\)-observable. In this case, the CNOT has the following effect on the qubits' expectation values between \(t_0-1\) and \(t_0\):
\begin{align} \label{eq:expectation.measurement}
\left\{
\begin{aligned}
\langle \hat{q}_a(t_0-1) \rangle_{\hat{\varrho}} & = (0, 0, 1) \\
\langle \hat{q}_b(t_0-1) \rangle_{\hat{\varrho}} & = (1, 0, 0)
\end{aligned}
\right\}
\overset{\text{CNOT}}{\longrightarrow}
\left\{
\begin{aligned}
\langle \hat{q}_a(t_0) \rangle_{\hat{\varrho}} & = (0, 0, 0) \\
\langle \hat{q}_b(t_0) \rangle_{\hat{\varrho}} & = (0, 0, 0)
\end{aligned}
\right\}
\end{align}

Evidently, the \(z\)-observables of both the control and the target qubits are non-sharp at time \(t_0\), so the measurer cannot have recorded a definite result. Instead, after the measurement, there are two instances of the measurer, one of which has registered `it was a \(1\)' and the other `it was a \(-1\)'. To make apparent that these two instances of the measurer exist, one has to consider the states of the measurer relative to the control qubit's \(z\)-observable having either of those values. To that end, I will define the following two projectors
\begin{align} \label{eq:projectors}
  \hat{\Pi}_{-1} (t) & \stackrel{\text{def}}{=}  \frac{1}{2} \bigl( \hat{1} - \hat{q}_{bz}(t) \bigr), \\
\hat{\Pi}_1 (t)  & \stackrel{\text{def}}{=}  \frac{1}{2} \bigl( \hat{1} + \hat{q}_{bz}(t) \bigr),
\end{align}
which satisfy an algebra that is characteristic of orthogonal projectors, i.e.
\begin{equation}\label{eq:projectoralgebra}
   \hat{\Pi}_{\alpha} (t) \hat{\Pi}_\beta (t)= \delta_{\alpha \beta} \hat{\Pi}_\alpha (t)  \qquad \qquad (\alpha, \beta\in \{1,-1\}).
\end{equation}
Together, these projectors form a projection valued measure (PVM) because they sum to unity, as can be readily verified from their expression in Eq.~\eqref{eq:projectors}. 

The expectation values of these projectors represent the probabilities of the \(z\)-observable of \(\mathcal{Q}_b\) having values \(1\) and \(-1\). These projectors are useful because they allow one to \textit{foliate} the descriptors of \(\mathcal{Q}_a\) into \textit{relative descriptors}, which in this case will represent the instances of \(\mathcal{Q}_a\) that recorded different measurement outcomes.

In particular, the descriptors of \(\mathcal{Q}_a\) relative to \(\hat{q}_{bz}(t_0)\) having value \(-1\) are defined as
\begin{align} \label{eq:relativedescriptors1}
&  \hat{\boldsymbol{q}}_{a,-1} (t)  \stackrel{\text{def}}{=} \hat{\Pi}_{-1}(t_0)  \hat{\boldsymbol{q}}_a (t).
\end{align}
These relative descriptors represent an instance of \(\mathcal{Q}_a\) that will be denoted as \(\mathcal{Q}_{a,-1}\). The relative descriptors are defined at a time \( t \geq t_0 \) relative to the value of \(\hat{q}_{bz}(t_0)\) because descriptors should depend only on systems within their own past light cone to ensure there is no action at a distance. Here, I assume that \(\hat{q}_{bz}(t_0)\) commutes with the components of \( \hat{\boldsymbol{q}}_a (t)\), as this is a requirement for a valid foliation into relative states. This requirement is certainly satisfied by \(\hat{q}_{bz}(t_0)\) and \( \hat{\boldsymbol{q}}_a (t_0)\). A sufficient (though not a necessary) condition for it to remain satisfied is that the qubits do not interact again. Similarly, there exists a second triple of descriptors of \(\mathcal{Q}_a\) that represent the value of the descriptors of that qubit relative to \(\hat{q}_{bz}(t_0)\) having the value \(1\), namely 
\begin{equation}\label{eq:relativedescriptors2}
 \hat{\boldsymbol{q}}_{a,1} (t)  \stackrel{\text{def}}{=} \hat{\Pi}_{1}(t_0)   \hat{\boldsymbol{q}}_a (t).
\end{equation}
Here, again, \(t \geq t_0\) . These relative descriptors represent an instance of \(\mathcal{Q}_a\) that will be denoted as \(\mathcal{Q}_{a,1}\). The additional subscripts \(1\) and \(-1\) on the left-hand sides of Eqs.~\eqref{eq:relativedescriptors1} and \eqref{eq:relativedescriptors2} are included to represent relative to which eigenvalue of \(\hat{q}_{bz}(t)\) the descriptors are being described. Also note that, since the projectors are assumed to commute with the descriptors \( \hat{\boldsymbol{q}}_a (t)\), the products in Eqs.~\eqref{eq:relativedescriptors1} and \eqref{eq:relativedescriptors2} are well-defined.

The projectors act in a way that is similar to scalars, so their effect in Eqs.~\eqref{eq:relativedescriptors1} and \eqref{eq:relativedescriptors2} is that the projectors multiply each of the components of the triple. Consequently, the algebra of the relative descriptors is different from that of the absolute ones in that the absolute descriptors adhere to the Pauli algebra relative to the unit \(\hat{1}\). For the relative descriptors, there is a different observable that plays the role of the unit, namely the projector \(\hat{\Pi}_{\theta}(t_0)\) with \(\theta \in \{ 1,-1\}\), and with respect to this projector, the relative descriptors do adhere to the Pauli algebra since
\begin{align}
\hat{q}_{ai,\theta}(t)   \hat{q}_{aj,\theta}(t) = \delta_{ij} \hat{\Pi}_{\theta}(t_0)  +i\epsilon_{ij}^{\ \ \ k} \hat{q}_{ak,\theta} (t) \qquad \qquad (i,j \in \{x,y,z\} ).
\end{align}
Thus, the projector \(\hat{\Pi}_{\theta}(t_0) \) has the role of the unit observable for each \(\hat{q}_{ai,\theta}(t)\); this is further corroborated by the fact that 
\begin{align}
\hat{q}_{ai,\theta}(t)  \hat{\Pi}_{\theta} (t_0)  = \hat{\Pi}_{\theta} (t_0) \hat{q}_{ai,\theta}(t) = \hat{q}_{ai,\theta}(t)  \qquad \qquad (i,j \in \{x,y,z\} ),
\end{align}
which is the same effect as the unit observable had in Eq.~\eqref{identity}.

To complete the foliation of the descriptors, one must include a \textit{relative expectation value}, defined as the expectation value conditioning on \( \hat{q}_{bz} (t_0) \) taking the value \(1\) or \(-1\). As such, let \(\theta \in \{1,-1\}\); then this relative expectation value is given by
\begin{align}
&  \langle \hat{q}_{ai,\theta}(t) \rangle_{\hat{\varrho},\theta}  \stackrel{\text{def}}{=} \frac{\langle \hat{q}_{ai,\theta}(t) \rangle_{\hat{\varrho}}  }{\langle  \hat{\Pi}_{\theta} (t_0)  \rangle_{\hat{\varrho}}}.
\end{align}
The expectation value of \(\langle  \hat{\Pi}_{\theta} (t_0)  \rangle_{\hat{\varrho}}\) is assumed to be non-zero for all \(\theta \in \{1,-1\}\), which is the second and final condition for this foliation into relative descriptors to be a valid one.

The relative expectation values of the relative descriptors at time \(t_0\) are 
\begin{align}
     \langle  \hat{\boldsymbol{q}}_{a,-1} (t_0) \rangle_{\hat{\varrho},{-1}} & = (0,0,-1),\\
     \langle  \hat{\boldsymbol{q}}_{a,1} (t_0) \rangle_{\hat{\varrho},{1}} & =  (0,0,1).
\end{align}
Notably, both relative \(z\)-observables are sharp since their variance with respect to the relative expectation values vanishes, i.e.
\begin{align}
    \langle  \hat{q}_{az, \theta}(t_0) \rangle_{\hat{\varrho},{\theta}} ^2 -  \langle  ( \hat{q}_{az, \theta}(t_0))^2 \rangle_{\hat{\varrho},{\theta}} =0,
\end{align}
where I have used the expression for the CNOT gate in Eq.~\eqref{eq:CNOT}, as well as the expectation values of the descriptors at time \(t_0-1\) and \(t_0 \), as expressed in Eq.~\eqref{eq:expectation.measurement}, and the fact that the qubits are in a product state relative to each other at time \(t_0-1\).

Evidently, the relative descriptor \(\hat{q}_{az, 1}(t_0)\) is sharp with value \(1\), just as  \(\hat{q}_{az, 1}(t_0)\) is sharp with value \(-1\), despite the absolute descriptor \(\hat{q}_{az}(t_0)\) being non-sharp. This is an important feature of the relative descriptors, as it helps explain the emergence of classicality from quantum systems. For instance, the macroscopic objects of our everyday experience, such as tables and chairs, appear to have observables with definite values – like definite positions, volume, magnetisation, and so forth – yet it is overwhelmingly likely that such systems are entangled with other systems due to the effects of decoherence, so these observables should be non-sharp. The appearance of sharpness is due to the fact that we perceive a system's relative observables, which can be sharp (or almost sharp) even when the absolute observables of a system aren't.

\subsection{The relative noumenal states}
The relative noumenal states are the pairs \( (\hat{\boldsymbol{q}}_{a,-1} (t),  \hat{\varrho})  \) and \( (\hat{\boldsymbol{q}}_{a,1} (t),\hat{\varrho} )\), which can be obtained from the absolute noumenal state through two \textit{relative noumenal projectors}
\begin{align}
    \pi_{1} \bigl( (\hat{\boldsymbol{q}}_{a} (t), \hat{\varrho} ) \bigr) & \stackrel{\text{def}}{=} (\hat{\boldsymbol{q}}_{a} (t) \hat{\Pi}_{1} (t_0), \hat{\varrho} ) =  (\hat{\boldsymbol{q}}_{a,1} (t),  \hat{\varrho}), \\
    \pi_{-1} \bigl( (\hat{\boldsymbol{q}}_{a} (t), \hat{\varrho} ) \bigr) & \stackrel{\text{def}}{=} (\hat{\boldsymbol{q}}_{a} (t) \hat{\Pi}_{-1} (t_0), \hat{\varrho} ) =  (\hat{\boldsymbol{q}}_{a,-1} (t),  \hat{\varrho}).
\end{align}
These relative noumenal projectors are simply a more formal way of expressing the foliation into relative states that has already been presented in Eqs.~\eqref{eq:relativedescriptors1} and \eqref{eq:relativedescriptors2}.

These relative noumenal states are separable and satisfy no-action-at-a-distance. Firstly, the two relative noumenal states can be recombined into the noumenal state of the absolute system, i.e. \( (\hat{\boldsymbol{q}}_{a} (t), \hat{\varrho} )\), through the addition of the relative descriptors. This suggests that one can define a \textit{relative noumenal product}, denoted \(\odot_R\), whose effect is to recombine the relative noumenal states back into the absolute one, as follows:
\begin{align}
    (\hat{\boldsymbol{q}}_{a,1} (t), \hat{\varrho} ) \odot_R  (\hat{\boldsymbol{q}}_{a,-1} (t), \hat{\varrho} ) & \stackrel{\text{def}}{=} (\hat{\boldsymbol{q}}_{a,1} (t)+\hat{\boldsymbol{q}}_{a,-1} (t), \hat{\varrho} ), \\
&= (\hat{\boldsymbol{q}}_{a} (t), \hat{\varrho} ).
\end{align}
Note that the relative descriptors sum back to the absolute ones because the projectors \(\hat{\Pi}_{1}(t_0)\) and  \(\hat{\Pi}_{-1}(t_0)\) constitute a PVM, implying they sum to unity. Since the relative descriptors can be recombined into the absolute descriptors, they satisfy the separability requirement expressed in Eq.~\eqref{eq:separability}. 

Similarly, the relative noumenal states satisfy the requirement that there is no action at a distance, as expressed in Eq.~\eqref{eq:localevolution}, because operations performed on each of the relative noumenal states separately can be combined into operations on the absolute state. To prove this, consider a unitary \(U\) that acts on  \( (\hat{\boldsymbol{q}}_{a} (t), \hat{\varrho} )\). For the foliation into relative noumenal states to be preserved by the dynamics, the gate \(U\) must commute with \(\hat{q}_{bz}(t_0)\). In that case, it can be decomposed into two parts, which I will call the \textit{relative unitaries}.
\begin{equation}
    U_{1}  \stackrel{\text{def}}{=}   U \hat{\Pi}_{1}(t_0), \qquad \qquad  U_{-1} \stackrel{\text{def}}{=}   U \hat{\Pi}_{-1}(t_0) .
\end{equation}
Here, \(U_{1}\) is a unitary that acts exclusively on the relative descriptors \(\hat{\boldsymbol{q}}_{a,1} (t) \), and likewise, \(U_{-1}\) acts exclusively on \(\hat{\boldsymbol{q}}_{a,-1} (t) \). It then follows from the fact the projectors sum to unity that
\begin{equation} \label{eq:sum.of.unitaries}
    U_{1} + U_{-1} = U,
\end{equation}
and moreover, after the unitary has been implemented, the relative descriptors still sum to the absolute one:
\begin{equation} \label{eq:unitary.decomposition}
   U^\dagger_{-1} \hat{\boldsymbol{q}}_{a,-1} (t)  U_{-1} +  U^\dagger_{1} \hat{\boldsymbol{q}}_{a,1} (t)  U_{1} = U^\dagger \hat{\boldsymbol{q}}_{a} (t)  U,
\end{equation}

We are now in the position to demonstrate that the relative noumenal states have no action at a distance. The effect of the unitaries \(U_{1}\) and \(U_{-1}\) on the relative noumenal states is defined as
\begin{align}
    U_{1} \star (\hat{\boldsymbol{q}}_{a,1} (t), \hat{\varrho} ) & \stackrel{\text{def}}{=}  (U_{1}^{\dagger} \hat{\boldsymbol{q}}_{a,1} (t) U_{1}, \hat{\varrho} ), \\
  U_{-1} \star (\hat{\boldsymbol{q}}_{a,1} (t), \hat{\varrho} ) & \stackrel{\text{def}}{=}  (U_{-1}^{\dagger} \hat{\boldsymbol{q}}_{a,-1} (t) U_{-1}, \hat{\varrho} ).
\end{align}
Using Eq.~\eqref{eq:unitary.decomposition}, one can readily verify that the relative noumenal states can still be recombined into the absolute one after the application of the unitaries on the relative noumenal states since
\begin{align} \label{eq:relative.evolution}
    \left ( U_{-1} \star (\hat{\boldsymbol{q}}_{a,-1} (t), \hat{\varrho}) \right ) \odot_R \left ( U_{1} \star (\hat{\boldsymbol{q}}_{a,1} (t), \hat{\varrho})) \right ) = U \star ( \hat{\boldsymbol{q}}_{a} (t), \hat{\varrho}).
\end{align}
No action at a distance, as expressed in Eq.~\eqref{eq:localevolution}, demands that there exists a so-called product of operations that allows one to take the operations that act on the relative noumenal states and combine them into operations on the absolute noumenal state. One can define a \textit{product of relative operations}, denoted \(\times_R\), as follows:
\begin{align} \label{eq:relative.product}
   (U_{-1} \times_R U_{1}) \star ( \hat{\boldsymbol{q}}_{a} (t), \hat{\varrho}) & \stackrel{\text{def}}{=} (U_{-1} + U_{1}) \star ( \hat{\boldsymbol{q}}_{a} (t), \hat{\varrho}).
\end{align}
 By combining the results of Eqs.~\eqref{eq:relative.evolution} and \eqref{eq:relative.product}, one can verify that Eq.~\eqref{eq:localevolution} is satisfied since
\begin{align}
    (U_{-1} \times_R U_{1}) \star ( \hat{\boldsymbol{q}}_{a} (t), \hat{\varrho}) & = (U_{-1} + U_{1}) \star ( \hat{\boldsymbol{q}}_{a} (t), \hat{\varrho}), \\
    & = U \star ( \hat{\boldsymbol{q}}_{a} (t), \hat{\varrho}), \\
    & = \left ( U_{-1} \star (\hat{\boldsymbol{q}}_{-1} (t), \hat{\varrho}) \right ) \odot_R \left ( U_{1} \star (\hat{\boldsymbol{q}}_{1} (t), \hat{\varrho})) \right ).
\end{align}
Here, to get from the first to the second line, I use the fact that the relative unitaries, \(U_{1}\) and \(U_{-1}\), sum back to the absolute one, \(U\), whenever the relative noumenal states are compatible. Evidently, \( (\hat{\boldsymbol{q}}_{a,1} (t), \hat{\varrho}) \) and \( (\hat{\boldsymbol{q}}_{a,-1} (t), \hat{\varrho}) \) satisfy the criteria imposed by local realism, implying they are bona fide noumenal states representing the two instances of \(\mathcal{Q}_a\). This completes the proof that the relative noumenal states comply with local realism.

It is noteworthy that I have only foliated one qubit, namely \(\mathcal{Q}_a\), into relative noumenal states without needing to simultaneously foliate \(\mathcal{Q}_b\), although one could. That it is possible to foliate one qubit without also foliating another qubit is a consequence of locality. In contrast, the Schrödinger picture relative states only allow one to foliate the entire multiverse, not just some of its subsystems. Hence, because the Schrödinger picture is not a separable description, branching in the Schrödinger picture is not local.\footnote{Blackshaw, Huggett, and Ladyman \cite{BlackshawEtAl2024} argue that branching is local in the Schrödinger picture, in the sense that interactions propagate locally under unitary dynamics. However, they acknowledge that quantum theory lacks a separable description, so a branch of the universal state vector encompasses all systems that exist. In contrast, I take the existence of a separable description to be a necessary condition for branching to count as local: when one system branches, the descriptions of remote systems should remain unaffected.}

\subsection{Local branching} \label{sec:local-branching}
A system, after becoming entangled with another (via a measurement or another entangling interaction), exists in multiple instances, each represented by a relative noumenal state. This branching of a noumenal state into relative noumenal states happens locally, which is what resolves the apparent non-locality in the EPR paradox. To see why, let us revisit the paradox. Alice and Bob each possess a qubit, \(\mathcal{Q}_A\) and \(\mathcal{Q}_B\), which are entangled. In the Schrödinger picture, Alice’s measurement of \(\mathcal{Q}_A\) seemed to instantaneously change Bob’s qubit, but this was due to the absence of a separable description that satisfies the no-action-at-a-distance principle. As has been shown, unitary quantum theory does admit such a local description.

How can the correlations between the outcomes of \(\mathcal{Q}_A\) and \(\mathcal{Q}_B\) be explained locally? In unitary quantum theory, Alice branches locally into two instances upon measuring \(\mathcal{Q}_A\), each unaware of the other. The same applies to Bob, who also branches locally upon measuring his qubit, with each instance observing a distinct outcome. Alice’s measurement has no effect on Bob’s noumenal state, and vice versa. Locality is preserved because only versions of Alice and Bob with matching outcomes can interact -- for example, only the Alice who observed a \(1\) can interact with the Bob who observed a \(1\). 

Brassard and Raymond-Robichaud describe this dramatically in \textit{Parallel Lives}~\cite{brassard2019parallel}: if the Bob who observed a \(1\) tried to meet the Alice who observed a \(-1\), they would pass through one another, entirely unaware of the other's existence. The correlations between Alice and Bob are thus established through local rules determining which versions can interact, without any action at a distance. Indeed, Bédard~\cite{bedard2024local} proves how this local description accounts for violations of Bell inequalities.

Notably, the correlations can only be established once Alice and Bob meet and become coincident in spacetime. Prior to that, there is no fact of the matter -- for either of them -- about the other’s outcome, or even whether a measurement has occurred. If Bob were to measure his qubit and later undo the measurement (which is possible in unitary quantum theory), then upon meeting Alice, neither would be aware of the result. Bob would have branched locally and undone this branching locally, without ever affecting the noumenal states of Alice. In the global account of branching, this is not possible: Alice would have temporarily been part of Bob's branches until they were undone. This is problematic from the standpoint of relativity.\footnote{Ney~\cite{Ney2024} describes branching as a global, frame-relative pseudo-process -- a change that appears to occur but lacks an underlying causal mechanism. Bacciagaluppi~\cite{Bacciagaluppi2025} similarly accepts global branching, but proposes a Schrödinger-picture construction that mitigates frame dependence and helps reconcile quantum measurement with special relativity. In contrast, I argue that branching should be understood as a local phenomenon: a separable description without action at a distance avoids nonlocal changes to remote systems and resolves frame dependence at its root.} 

In both special and general relativity, simultaneity is not absolute: there is no fixed order to spacelike-separated events. In some frames, event \(X\) precedes spacelike-separated event \(Y\); in others, the order is reversed. If Bob performs his measurement at event \(X\), spacelike-separated from Alice's event \(Y\), then in some frames the measurements are simultaneous, in others sequential, depending on the frame.

If Bob's measurement at \(X\) were to instantly and globally branch the multiverse (including Alice's system at \(Y\)), then in some frames this branching at \(Y\) would result from a future event, implying backward causation. The local account avoids this: a branching event initially affects only the measured system, and its effects propagate outward at no faster than light speed. As a result, only the region within the future light cone of \(X\) could be part of the same branching structure. Put differently, in the local Everettian view, branching occurs in bubbles of space, and nothing outside such a bubble is part of any branch within it.

\section{EPR in hindsight}
In light of the local theory provided here, the concerns of  Einstein, Rosen and Podolsky become more apparent. They were right that quantum theory, in the form it existed in their day, did not satisfy local realism. The reasons for this are that their understanding of quantum theory contained the collapse postulate, which produces action at a distance, and they expressed the theory in the Schrödinger picture, which is not a separable description. Thus,  Einstein, Rosen and Podolsky were right to raise concerns about locality and to conclude that quantum theory was not `complete'. The `complete' version of quantum theory that Einstein, Rosen and Podolsky desired was only achieved by formulating Everett's relative state construction in the Heisenberg picture – I believe that the local account of the quantum theory presented in this paper, which is a summary of the work of multiple researchers labouring over multiple decades, is what Einstein, Rosen and Podolsky had hoped for.

\newpage

\section*{Acknowledgements}
The author thanks David Deutsch, Charles Alexandre Bédard, Gilles Brassard, Eric Marcus, and Max Velthoven for their valuable feedback on earlier drafts of this work, which greatly improved its quality. He also expresses gratitude to Paul Raymond-Robichaud for helpful discussions and to Gilles Brassard for enabling this project. This research was supported by the Québec-Ontario Consortium on Quantum Protocols (QUORUM) and by Conjecture Institute.

\appendix 

\section*{Appendix}

\section{Local and non-local descriptions} \label{sec:derivation}

Theories that are known to be local often admit non-local descriptions. A theory should, thus, be considered local if it admits at least one local description. Deutsch \cite{deutsch2012vindication} illustrates this point through the example of a classical system of \(N\) Newtonian billiard balls. This system is local because the noumenal states of each ball consist of its position and momentum, and the balls interact through contact only. One can transition from this local description to a non-local one by performing a transformation that mixes some of the positions of each ball. If the positions of the \(N\) balls are \(x_1(t), \dots,x_N(t) \), one may define 
\begin{equation} 
x'_j(t) = \sum_k V_{j}^{\ k} x_k(t) .
\end{equation}

If the matrix \(V_{j}^{\ k}\) has an inverse, the new variables \(x'_1(t), \dots, x'_N(t) \) contain the same information as the original positions \(x_1(t), \dots,x_N(t) \), so both will yield the same predictions about the outcomes of measurements. However, if one were to interpret the \(x'_a(t) \) as describing the state of ball \(a\), then in general, this description would be non-local because \(x'_a(t) \) generically depends on the positions of the other balls in the system. Hence, operations performed on those balls will typically affect \(x'_a(t) \), violating the assumption that there is no action at a distance.

This situation is analogous to the relationship between the Heisenberg and Schrödinger pictures in quantum mechanics. These two pictures are related by a transformation that preserves the theory's predictions, yet the transformation connects a local description to a non-local one.

\section{EPR in terms of noumenal states} \label{sec:noumenal-EPR}
Let us revisit the EPR paradox and recast it in terms of noumenal states. I will present a proof by contradiction showing that no separable description in terms of the state vector can satisfy the no-action-at-a-distance principle.

Consider again the entangled state \(\ket{\Psi}_{AB}\) of Alice and Bob’s joint system, as expressed in Eq.~\eqref{eq:EPR-pair}. Measuring this state in the computational basis yields one of two possible outcomes. By measuring in the computational basis, the state could collapse to \(\ket{0}_A \ket{0}_B\) or \(\ket{1}_A \ket{1}_B\). We shall compare these possible measurement outcomes.

Let us assume that there exists a separable description for each of these possible post-measurement states, implying that
\begin{align}
\ket{1}_A \ket{1}_B = N'_A \odot N'_B, \\
\ket{0}_A \ket{0}_B = N''_A \odot N''_B,
\end{align}
where \( N'_A\) and \(N''_A\) are noumenal states of Alice's qubit, and \( N'_B\) and \(N''_B\) are noumenal states of Bob's qubit. Since Alice's measurement should only have affected Alice's qubit, we can conclude the noumenal state of Bob's qubit should be unaffected in either case, i.e. \(N'_B = N''_B\). We can also see that the states are related by a Hadamard gate performed on each qubit since
\begin{align}
\ket{1}_A \ket{1}_B & = (X\otimes X) \ket{0}_A \ket{0}_B,
\end{align}
where
\begin{align}
X = \begin{pmatrix}
0 & 1 \\
1 & 0 \end{pmatrix}.
\end{align}
It follows from the no-action-at-a-distance principle that
\begin{align}
 (X\otimes X) \star (N'_A \odot N'_B) = (X \star N'_A ) \odot (X \star N'_B) = N''_A \odot N''_B, 
\end{align}
which implies that \(X \star N'_A = N''_A\) and \(X \star N'_B = N''_B\). But now we can conclude that
\begin{align}
\ket{1}_A \ket{1}_B &= (X \star N'_A ) \odot (X \star N'_B), \\
&= (X \star N'_A ) \odot ( N'_B),\\
&= (X\otimes I_B) \star (  N'_A  \odot  N'_B), \\
&= (X \otimes I_B) \ket{0}_A \ket{0}_B, \\
&= \ket{1}_A \ket{0}_B.
\end{align}
Here, \(I_B\) is the unit operator of \(\mathcal{Q}_B\). Thus, we have derived a contradiction, implying that a separable description cannot satisfy the no-action-at-a-distance principle if quantum theory is formulated in the Schrödinger picture and the collapse postulate is assumed. A similar proof can be found in Ref.~\cite{raymond2021local}.

\bibliographystyle{unsrt}
\bibliography{mybib}

\end{document}